\documentstyle[12pt]{article}
\def\ba{\begin{eqnarray}}
\def\ea{\end{eqnarray}}

\def\lb{\label}
\def\be{\begin{equation}}
\def\ee{\end{equation}}

\begin{document}

\title{A K\"ahler-Einstein inspired anzatz for $Spin(7)$ holonomy metrics and its solution}
\author{O.P.Santillan \thanks{firenzecita@hotmail.com}}
\date{}
\maketitle

\begin{abstract}

 We construct propose an anzatz for $Spin(7)$ metrics as an $R$ bundle
over closed $G_2$ structures. These $G_2$ structures are $R^3$
bundles over 4-dimensional compact quaternion K\"ahler spaces. The
inspiration for the anzatz metric comes from the Bryant-Salamon
construction of $G_2$ holonomy metrics and from the fact that the
twistor space of any compact quaternion K\"ahler space is
K\"ahler-Einstein. The reduction of the holonomy to a subgroup of
$Spin(7)$ gives non linear system relating three unknown functions
of one variable. We obtain a particular solution and we find that
the resulting metric is a Calabi-Yau cone over an Einstein-Sassaki
manifold which means that the holonomy is reduced to $SU(4)\subset
Spin(7)$. Another coordinate change show us that our metrics are
hyperkahler cones known as Swann bundles, thus the holonomy is
reduced to $Sp(2)\subset SU(4)\subset Spin(7)$ and the cone is
tri-Sassakian. We revert our argument and state that the Swann
bundle define a closed $G_2$ structure by reduction along an
isometry. We calculate the torsion classes for such structure
explicitly.

\end{abstract}

\section{Introduction}

There exist a growing interest in the construction of $Spin(7)$
holonomy metrics due to their application in supergravity
compactification preserving certain amount of supersymmetry
\cite{Spin10}-\cite{Spin20}. The present work is concerned with this
task and from the analysis performed here it is obtained the
following proposition.
\\

{\bf Proposition }{ \it Let us consider a compact quaternion
K\"ahler space $M$ in $d=4$ with metric $g_q$ and with cosmological
constant $\Lambda$ normalized to $3$. For any of such metrics there
always exist a basis $e^a$ such that $g_q=\delta_{ab}e^a\otimes e^b$
for which the $Sp(1)$ part of the spin connection $\omega^{a}_{-}$
and the negative oriented K\"ahler triplet $\overline{J}_i$ defined
by
$$
\omega^{a}_{-}=\omega^a_{0}- \epsilon_{abc}\omega^b_c,\qquad
\overline{J}_1=e^1\wedge e^2-e^3\wedge e^4,
$$
$$
\overline{J}_2=e^1\wedge e^3-e^4\wedge e^2\qquad
\overline{J}_3=e^1\wedge e^4-e^2\wedge e^3,
$$
satisfy the relations \be\lb{relat}
d\omega_{-}^i+\epsilon_{ijk}\omega_{-}^j \wedge \omega_{-}^k=
\overline{J}_i, \qquad
d\overline{J}^i=\epsilon_{ijk}\overline{J}^{j}\wedge\omega_{-}^{k}.
\ee Let $\tau$ and $u_i$ be four new coordinates, $u=\sqrt{u^iu^i}$,
$\alpha_i=du^i+\epsilon^{ijk}\omega_{-}^j u^k$ and $H$ a
$\tau$-independent one form. Then the 8-dimensional metric
\be\lb{eyo}
g_8=\frac{(dt+H)^2}{e^{\frac{3}{2}h}}+e^{2f+\frac{1}{2}h}
\alpha_i\alpha_i+e^{2g+\frac{1}{2}h} g_q \ee together with the four
form
$$
\Phi_4=(dt+H)\wedge\bigg( e^{3f}\alpha_1\wedge \alpha_2\wedge
\alpha_3+e^{f+2g}\alpha_i\wedge \overline{J}_i
\bigg)+e^{2(f+g)+h}\frac{\epsilon_{ijk}}{2}\alpha_i\wedge
\alpha_j\wedge \overline{J}_k+e^{4g+h}e_1\wedge e_2\wedge e_3\wedge
e_4,
$$
constitute an $Spin(7)$ structure preserved by the Killing vector
$\partial_{\tau}$. Moreover if $f$, $g$ and $h$ are functions of $u$
related by
$$
u e^{3f}=(e^{f+2g})', \qquad
\lambda(e^{3f}-\frac{e^{f+2g}}{u^2})=(e^{2(f+g)+h})',
$$
$$4 ue^{2(f+g)+h}-2\lambda e^{f+2g}=(e^{4g+h})', $$ and the 1-form
$H$ satisfy \be\lb{mur}
dH=-\widetilde{u}_i\overline{J}_i+\frac{\epsilon_{ijk}}{2}\widetilde{u}_i\theta_j\wedge
\theta_k, \ee being
$\theta_i=d(\widetilde{u}^i)+\epsilon^{ijk}\omega_{-}^j
\widetilde{u}^k$ and $\widetilde{u}^i=u^i/u$, then $\Phi_4$ will be
closed and therefore, the holonomy of the metric (\ref{eyo}) will be
included in $Spin(7)$.}
\\

   We will show below that the integrability condition for
(\ref{mur}), namely
$$
d(\widetilde{u}_i\overline{J}_i-\frac{\epsilon_{ijk}}{2}\widetilde{u}_i\theta_j\wedge
\theta_k)=0
$$
is always satisfied due to the fact that the twistor space $Z$ of
any compact quaternion K\"ahler space $M$ carries a
K\"ahler-Einstein metric of positive scalar curvature \cite{Salomon}
and the left hand side of (\ref{mur}) is the K\"ahler form for such
metric. Also, the closure of $\Phi_4$ follows directly from the
formulas (\ref{use}) given below. By construction, the vector field
$\partial_{\tau}$ is Killing and if the quaternion K\"ahler basis
possess an isometry group $G$ which preserves the forms
$\omega_{-}^i$, then $G$ will be an isometry of $g_8$.  The
construction of $Spin(7)$ holonomy metrics that follows from the
proposition can also be applied to quaternion K\"ahler orbifolds.

\section{Closed $G_2$ and $Spin(7)$ structures and K\"ahler-Einstein metrics}

\textit{Quaternion K\"ahler spaces in brief}
\\

   A key ingredient in order to construct the metrics (\ref{eyo}) are quaternion
K\"ahler manifolds and so, it is convenient to give a brief
description of their properties. By definition, a quaternion
K\"ahler space $M$ is an euclidean $4n$ dimensional space with
holonomy group $\Gamma$ included into the Lie group $Sp(n)\times
Sp(1)\subset SO(4n)$ \cite{Berger}-\cite{Ishihara}. This affirmation
is non trivial if $D>4$, but in $D=4$ we have the well known
isomorphism $Sp(1)\times Sp(1)\simeq SU(2)_L\times SU(2)_R \simeq
SO(4)$ and so to state that $\Gamma\subseteq Sp(1)\times Sp(1)$ is
the same that to state that $\Gamma\subseteq SO(4)$. The last
affirmation is trivially satisfied for any oriented space and gives
almost no restrictions about the space, therefore the definition of
quaternion K\"ahler spaces should be modified in $d=4$. Their main
properties are the following.
\\

- There exists three automorphism $J^i$ ($i=1$ ,$2$, $3$) of the
tangent space $TM_x$ at a given point $x$ with multiplication rule
$J^{i} \cdot J^{j} = -\delta_{ij} + \epsilon_{ijk}J^{k}$, and for
which the metric $g_q$ is quaternion hermitian, that is
\be\lb{hermoso} g_q(X,Y)=g(J^i X, J^i Y), \ee being $X$ and $Y$
arbitrary vector fields.
\\

- The automorphisms $J^i$ satisfy the fundamental relation
\be\lb{rela2} \nabla_{X}J^{i}=\epsilon_{ijk}J^{j}\omega_{-}^{k}, \ee
with $\nabla_{X}$ the Levi-Civita connection of $M$ and
$\omega_{-}^{i}$ its $Sp(1)$ part. As a consequence of hermiticity
of $g$, the tensor $\overline{J}^{i}_{ab}=(J^{i})_{a}^{c}g_{cb}$ is
antisymmetric, and the associated 2-form
$$
\overline{J}^i=\overline{J}^{i}_{ab} e^a \wedge e^b
$$
satisfies \be\lb{basta}
d\overline{J}^i=\epsilon_{ijk}\overline{J}^{j}\wedge\omega_{-}^{k},
\ee being $d$ the usual exterior derivative.
\\

- Corresponding to the $Sp(1)$ connection we can define the 2-form
$$
F^i=d\omega_{-}^i+\epsilon_{ijk}\omega_{-}^j \wedge \omega_{-}^k.
$$
Then for a quaternion K\"ahler manifold \be\lb{lamas}
R^i_{-}=2n\kappa \overline{J}^i, \ee \be\lb{rela} F^i=\kappa
\overline{J}^i, \ee being $\Lambda$ certain constant and $\kappa$
the scalar curvature. The tensor $R^a_{-}$ is the $Sp(1)$ part of
the curvature. The last two conditions implies that $g$ is Einstein
with non zero cosmological constant, i.e, $R_{ij}=3\kappa
(g_{q})_{ij}$ being $R_{ij}$ the Ricci tensor constructed from
$g_q$. Notice that (\ref{rela}) is equivalent to (\ref{relat}) if we
choose the normalization $\kappa=1$.
\\

- For any quaternion K\"ahler space the $(0,4)$ and $(2,2)$ tensors
$$
\Theta=\overline{J}^1 \wedge \overline{J}^1 + \overline{J}^2 \wedge
\overline{J}^2 + \overline{J}^3 \wedge \overline{J}^3,
$$
$$
\Xi= J^1 \otimes J^1 + J^2 \otimes J^2 + J^3 \otimes J^3
$$
are globally defined and covariantly constant with respect to the
usual Levi Civita connection.
\\

- Any quaternion K\"ahler space is orientable.
\\

- In four dimensions  the K\"ahler triplet $\overline{J}_2$ and the
one forms $\omega^{a}_{-}$ are
$$
\omega^{a}_{-}=\omega^a_{0}- \epsilon_{abc}\omega^b_c,\qquad
\overline{J}_1=e^1\wedge e^2-e^3\wedge e^4,
$$
$$
\overline{J}_2=e^1\wedge e^3-e^4\wedge e^2\qquad
\overline{J}_3=e^1\wedge e^4-e^2\wedge e^3.
$$
In this dimension quaternion K\"ahler spaces are defined by the
conditions (\ref{rela}) and (\ref{lamas}). This definition is
equivalent to state that quaternion K\"ahler spaces are Einstein and
with self-dual Weyl tensor.
\\

\textit{The twistor space of a quaternion K\"ahler space}
\\

  Another very important property about compact quaternion K\"ahler spaces
is that its twistor space is \emph{K\"ahler-Einstein}. In order to
define the twistor space let us note that any linear combination of
the form $J=\widetilde{u}_i J_i$ is an almost complex structure on
$M$, and  the metric $g_q$ is hermitic with respect to it. Here we
have defined the scalar fields $\widetilde{u}^i=u^i/u$ and it is
evident that they are constrained by the condition $\widetilde{u}^i
\widetilde{u}^i=1$. This means that the bundle of almost complex
structures over $M$ is parameterized by points on the two sphere
$S^2$. This bundle is known as the twistor space $Z$ of $M$. The
space $Z$ is endowed with the metric \be\lb{kahlo} g_6=\theta_i
\theta_i + g_q,\ee where we have defined
$$
\theta_i=d(\widetilde{u}^i)+\epsilon^{ijk}\omega_{-}^j
\widetilde{u}^k.
$$
The metric (\ref{kahlo}) is six dimensional due to the constraint
$\widetilde{u}^i \widetilde{u}^i=1$. Corresponding to this metric we
have the K\"ahler two form \be\lb{two} \overline{J}=
\widetilde{u}_i\overline{J}_i-\frac{\epsilon_{ijk}}{2}\widetilde{u}_i\theta_j\wedge
\theta_k. \ee It has been proved in \cite{Salomon} that $J$ is
integrable and $\overline{J}$ is closed (see also \cite{Lolo}),
therefore $J$ is truly a complex structure and $g_6$ is
\emph{K\"ahler}. The calculation of the Ricci tensor of $g_6$ shows
that it is also Einstein, therefore the space $Z$ is
\emph{K\"ahler-Einstein}. We are using the normalization $\kappa=1$
here, for other normalization certain coefficients must be included
in (\ref{two}). Let us introduce the covariant derivative
\be\lb{alf} \alpha_i=du^i+\epsilon^{ijk}\omega_{-}^j u^k, \ee which
is related to $\theta_i$ by
$$
\theta^i=\frac{\alpha_i}{u}-\frac{u_i du}{u^2}.
$$
With the help of this relation and the definition (\ref{two}) it
follows that \be\lb{two2} \overline{J}=\frac{u_i}{u}\overline{J}_i
-\frac{\epsilon_{ijk}}{2}u_i\frac{\alpha_j\wedge
\alpha_k}{u^3}-\epsilon_{ijk}u_iu_j\frac{\alpha_k\wedge du}{u^4}\ee
$$ =\frac{u_i}{u}\overline{J}_i
-\frac{\epsilon_{ijk}}{2}u_i\frac{\alpha_j\wedge \alpha_k}{u^3}.
$$ The last expression will be needed in the following, although it is completely equivalent
to (\ref{two}). Also, the following formulae
$$
\overline{J}_i\wedge \overline{J}_j=-2\delta_{ij}e_1 \wedge
e_2\wedge e_3\wedge e_4,
$$
$$
d\alpha_i=\epsilon_{ijk}(u_j\overline{J}_k+\alpha_k\wedge
\omega_j^-),
$$
$$
d(u^iu^i)=d(u^2)=2u du=2u^i\alpha_i,
$$
\be\lb{use} d(\epsilon_{ijk}\alpha_i\wedge \alpha_j\wedge\alpha_k)=-
u du\wedge \alpha_i\wedge \overline{J}_i, \ee
$$
d(\alpha_i\wedge \overline{J}_i)=0,
$$
$$
d(e^{3f})\wedge\alpha_1\wedge
\alpha_2\wedge\alpha_3=(e^{3f})'du\wedge\alpha_1\wedge
\alpha_2\wedge\alpha_3=0,
$$
relating $\overline{J}_i$ and $\alpha_i$ will be useful for our
purposes. \footnote{A more complete account of formulae can be
found, for instance, in \cite{bernie}.} For instance, the closure of
(\ref{two2}) is a direct consequence of the second (\ref{use})
together with (\ref{basta}).
\\

\textit{A proof of the proposition}
\\

   Let us go back to our task of constructing the metric (\ref{eyo}).
Our starting point is an eight-dimensional metric anzatz of the form
\be\lb{anzatz}
g_8=\frac{(dt+H)^2}{e^{\frac{3}{2}h}}+e^{\frac{1}{2}h}g_7, \ee being
$g_7$ a metric over a 7-manifold $Y$ and $h$ a function over $Y$.
Neither the one form $H$ nor the function $h$ depends on $t$,
therefore the vector field $\partial_t$ is, by construction,
Killing. Associated to the metric (\ref{anzatz}) we can construct
the octonionic $4$-form
 \be\lb{ol}
\Phi_4=(dt+H)\wedge \Phi+e^{h}\ast\Phi, \ee being $\Phi$ a $G_2$
invariant three form corresponding to the metric $g_7$ and $\ast
\Phi$ its dual. The precise form for $\Phi$ will be found below. If
we impose the condition $d\Phi_4=0$ then the metric $g_8$ will have
$Spin(7)$ holonomy. \footnote{The converse of this affirmation is
not true, that is, the holonomy group of $g_8$ could be $Spin(7)$
and (\ref{ol}) could be not closed. The form (\ref{ol}) is preserved
by the Killing vector, but there could exist cases for which this
simplifying condition do not hold and the holonomy is still
$Spin(7)$. In such cases there will exist another closed 4-form
$\Phi_4$ which is not preserved by $\partial_t$.} We will suppose
that the seven dimensional metric $g_7$ is of the form \be\lb{anz}
g_7=e^{2f} \alpha_i\alpha_i+e^{2g} g_q, \ee being $g_q$ a quaternion
K\"ahler metric in $d=4$ and $\alpha_i$ defined in (\ref{alf}). The
functions $f$, $g$ will depend only on the "radius"
$u=\sqrt{u^iu^i}$. The form (\ref{anz}) for the 7-metric is well
known and is inspired in the Bryant-Salamon construction for $G_2$
holonomy metrics \cite{Bryant}. For a metric with $G_2$ holonomy we
have $d\Phi=d\ast \Phi=0$ but we will not suppose that the holonomy
of (\ref{anz}) is $G_2$, as in the Bryant-Salamon case. Instead we
will consider 7-spaces for which the form $\Phi$ is closed but not
co-closed. These are known as \emph{closed $G_2$ structures}
\cite{Chiozzi}-\cite{Spin12}. In this case the $Spin(7)$ holonomy
condition $d\Phi_4=0$ for (\ref{ol}) will reduce to \be\lb{imita}
dH\wedge \Phi=-d(e^h\ast \Phi). \ee We will find below a suitable
$H$ and $g_7$ for which (\ref{imita}) gets simplified even more.  It
seems reasonable for us to choose $H$ in (\ref{ol}) such that
\be\lb{h} dH=-\lambda\overline{J}. \ee The reason for this election
is that the integrability condition $d\overline{J}=0$ will be
automatically satisfied because, as we have seen above, the two form
$\overline{J}$ is the K\"ahler form of a K\"ahler-Einstein metric.
Here $\lambda$ is a parameter, and the minus sign was introduced by
convenience. Also, by selecting the basis
$$
\widetilde{e}_{i}=e^f\alpha_i,\qquad i=1,2,3\qquad
\widetilde{e}_{\alpha}= e^g e_{\alpha} \qquad \alpha=1,2,3,4,
$$
for (\ref{anz}), we can construct the $G_2$ invariant three form
 \be\lb{tre}
\Phi=c_{abc}\widetilde{e}^a\wedge \widetilde{e}^b\wedge
\widetilde{e}^c= e^{3f}\alpha_1\wedge \alpha_2\wedge
\alpha_3+e^{f+2g}\alpha_i\wedge \overline{J}_i \ee and its dual
\be\lb{cuat}
\ast\Phi=e^{2(f+g)}\frac{\epsilon_{ijk}}{2}\alpha_i\wedge
\alpha_j\wedge \overline{J}_k+e^{4g}e_1\wedge e_2\wedge e_3\wedge
e_4. \ee Here $e^{\alpha}$ is a basis for the quaternion K\"ahler
metric $g_q$. As we stated above, we will consider 7-spaces for
which the form $\Phi$ is closed but not co-closed. In other words we
will have that $d\Phi=0$ but $d\ast\Phi\neq 0$. We also suppose that
$f$, $g$ and $h$ are functions of the radius $u$ only. By using
(\ref{use}) it follows that the closure condition $d\Phi=0$ for
(\ref{tre}) leads to the equation \be\lb{closo}  u
e^{3f}=(e^{f+2g})', \ee where the tilde implies the derivation with
respect to $u$. This is one of the equations that we need.

   By another side, with the election (\ref{h}) for $H$ and using that $d\Phi=0$ it
follows from (\ref{ol}) that \be\lb{cla} d\Phi_4=dH\wedge
\Phi+d(e^{h}\ast\Phi)=-\lambda\overline{J}\wedge\Phi+d(e^{h}\ast\Phi)
\ee and therefore the $Spin(7)$ holonomy condition $d\Phi_4=0$ is
equivalent to \be\lb{clad}
\lambda\overline{J}\wedge\Phi=d(e^{h}\ast\Phi). \ee From (\ref{tre})
and (\ref{two2}) we see that the left side of (\ref{clad}) is
 \be\lb{left}
\lambda\overline{J}\wedge\Phi=\lambda(\frac{e^{3f}}{u}
-\frac{e^{f+2g}}{u^3})u^i\overline{J}^i\wedge\alpha_1\wedge
\alpha_2\wedge \alpha_3-2\lambda \frac{e^{f+2g}}{u}u^i\alpha_i\wedge
e_1\wedge e_2\wedge e_3\wedge e_4. \ee By using the formula
$u^i\alpha_i=u du$ and that
$$
u^i\overline{J}^i\wedge\alpha_1\wedge \alpha_2\wedge \alpha_3=u
du\wedge \frac{\epsilon_{ijk}}{2}\alpha_i\wedge
\alpha_j\wedge\overline{J}^k,
$$
we can reexpress (\ref{left}) as \be\lb{alr}
\lambda\overline{J}\wedge\Phi=\lambda(e^{3f}-\frac{e^{f+2g}}{u^2})
du\wedge \frac{\epsilon_{ijk}}{2}\alpha_i\wedge
\alpha_j\wedge\overline{J}^k-2\lambda e^{f+2g} du\wedge e_1\wedge
e_2\wedge e_3\wedge e_4. \ee By another side, the right side of the
equation (\ref{clad}) is found directly from (\ref{cuat}) and
(\ref{use}), the result is
 \be\lb{right}
d(e^{h}\ast\Phi)=(e^{2(f+g)+h})'du\wedge
\frac{\epsilon_{ijk}}{2}\alpha_i\wedge
\alpha_j\wedge\overline{J}^k+\bigg((e^{4g+h})'-4
ue^{2(f+g)+h}\bigg)du\wedge e_1\wedge e_2\wedge e_3\wedge e_4. \ee
By equating (\ref{right}) with (\ref{alr}) and taking into account
the closure condition (\ref{closo}) we obtain the following
differential system
$$
u e^{3f}=(e^{f+2g})', \qquad
\lambda(e^{3f}-\frac{e^{f+2g}}{u^2})=(e^{2(f+g)+h})',
$$
\be\lb{sistema} 4 ue^{2(f+g)+h}-2\lambda e^{f+2g}=(e^{4g+h})'. \ee
From this system of equations we obtain the proposition stated
above.

\subsection{A particular solution: the Swann bundle}

  If we were able to solve the system (\ref{sistema}) then we will
obtain a family of $Spin(7)$ holonomy metrics based on arbitrary
quaternion K\"ahler spaces. We do not know its general solution, but
we have found a particular one. It is not difficult to check that
indeed \be\lb{ck} e^f=u^{-1/3},\qquad e^g=u^{2/3},\qquad e^h=\lambda
u^{-2/3}, \ee is a solution of (\ref{sistema}). By introducing it
into the expression (\ref{anzatz}) and by defining the variable
$\tau=t/\lambda$ and rescaling by $g_8\to\lambda^{-1}g_8$ we obtain
the following metric \be\lb{eyo}
g_8=u(d\tau+H)^2+\frac{\alpha_1^2+\alpha_2^2+\alpha_3^2}{u} + u g_q,
\ee and the corresponding closed four form is
$$
\Phi_4=(d\tau+H)\wedge\bigg( \frac{\alpha_1\wedge \alpha_2\wedge
\alpha_3}{u}+u\alpha_i\wedge \overline{J}_i
\bigg)+\frac{\epsilon_{ijk}}{2}\alpha_i\wedge \alpha_j\wedge
\overline{J}_k+u^2 e_1 \wedge e_2\wedge e_3\wedge e_4.
$$
The closure of this form follows directly from formulas (\ref{use}).
An inspection of this metric shows that it is a cone over an
Einstein-Sassaki metric, thus the holonomy is in $SU(4)\in G_2$. In
order to see this we need to show the orthogonality condition
$\widetilde{u}_i\theta_i=0$, which is a consequence of the following
calculation
$$
\widetilde{u}_i\theta_i=\widetilde{u}_i
d\widetilde{u}_i+\epsilon^{ijk}\widetilde{u}^i\omega_{-}^j
\widetilde{u}^k=\widetilde{u}_i d\widetilde{u}_i=
d(\widetilde{u}_i\widetilde{u}_i)=0,
$$
we have used $\widetilde{u}_i\widetilde{u}_i=1$ in the last line. We
also have that
$$
u\theta^i+\frac{u_i du}{u}=\alpha_i,
$$
and by inserting this expression in (\ref{eyo}), applying after the
orthogonality condition and by defining the new radius $r=2u^{1/2}$
gives the following conical form of the metric
\be\lb{cono} g_8=dr^2+r^2g_7, \ee being $g_7$ given by \be\lb{owner}
g_7=(d\tau+H)^2+g_6=(d\tau+H)^2+\theta_i \theta_i + g_q. \ee We have
seen that the six dimensional metric $g_6$ is K\"ahler-Einstein and
therefore $g_7$ is Einstein-Sassaki (see the lectures \cite{Galicki}
and references therein). Any cone over an Einstein-Sassaki space is
Calabi-Yau and therefore its holonomy is in $SU(4)\subset Spin(7)$.

   But more information about these metrics can be found by
finding explicitly the one form $H$, which is defined by
$dH=\overline{J}$. In order to solve $dH=\overline{J}$ we need to
simplify the expression (\ref{two}). Let us remind that
$$
\overline{J}=
\widetilde{u}_i\overline{J}_i-\frac{\epsilon_{ijk}}{2}\widetilde{u}_i\theta_j\wedge
\theta_k
$$
and that $\theta_i=d(\widetilde{u}^i)+\epsilon^{ijk}\omega_{-}^j
\widetilde{u}^k$. The orthogonality condition $\widetilde{u}_i\theta_i=0$
is equivalent to
$$
\theta_3=-\frac{(\widetilde{u}_1\theta_1+\widetilde{u}_2\theta_2)}{\widetilde{u}_3}.
$$
From the last relation it follows that
$$
\frac{\epsilon_{ijk}}{2}\widetilde{u}_i\theta_j\wedge
\theta_k=\frac{\theta_1\wedge \theta_2}{\widetilde{u}_3}.
$$
After certain calculation we obtain
$$
\frac{\theta_1\wedge \theta_2}{\widetilde{u}_3}
=\frac{d\widetilde{u}_1\wedge
d\widetilde{u}_2}{\widetilde{u}_3}-d\widetilde{u}_i\wedge
\omega_{-}^i+\frac{\epsilon^{ijk}}{2}\widetilde{u}_i\omega_{-}^j\wedge
\omega_{-}^k.
$$
Therefore
\be\lb{mam}
\frac{\epsilon_{ijk}}{2}\widetilde{u}_i\theta_j\wedge
\theta_k=\frac{d\widetilde{u}_1\wedge
d\widetilde{u}_2}{\widetilde{u}_3}-d\widetilde{u}_i\wedge
\omega_{-}^i+\frac{\epsilon^{ijk}}{2}\widetilde{u}_i\omega_{-}^j\wedge
\omega_{-}^k.
\ee
By another side we have the fundamental relation for quaternion Kahler manifolds, which is
\be\lb{mamsa}
\widetilde{J}_i=d\omega_{-}^i+\frac{\epsilon^{ijk}}{2}\omega_{-}^j\wedge
\omega_{-}^k.
\ee
Inserting expressions (\ref{mam}) and (\ref{mamsa}) into (\ref{two}) give us a
remarkably simple expression for $\overline{J}$, namely
\be\lb{simple}
\overline{J}=d(\widetilde{u}_i\omega_{-}^i)-\frac{d\widetilde{u}_1\wedge
d\widetilde{u}_2}{\widetilde{u}_3} \ee By parameterizing the
coordinates $\widetilde{u}_i$ in the spherical form
$$
\widetilde{u}_1=\cos\theta,\qquad
\widetilde{u}_2=\sin\theta \cos\varphi,\qquad
\widetilde{u}_3=\sin\theta \sin\varphi,\qquad
$$
we find out that
$$
\frac{d\widetilde{u}_1\wedge
d\widetilde{u}_2}{\widetilde{u}_3}=-d\varphi\wedge d\cos\theta.
$$
With the help of the last expression can reexpress (\ref{simple}) as
$$
\overline{J}=d(\widetilde{u}_i\omega_{-}^i)-d\varphi\wedge
d\cos\theta,
$$
from where it follows directly that the form $H$ such that
$dH=\overline{J}$ is given by \be\lb{simplon}
H=u_i\omega_{-}^i+\cos\theta d\varphi, \ee up to a total
differential term. By introducing the expression (\ref{simplon})
into (\ref{owner}) we find directly the following expression for the Einstein-Sasaki
metric
$$
g_7=(d\tau+\cos\theta d\varphi+\cos\theta\omega_{-}^1+\sin\theta \cos\varphi\omega_{-}^2
+\sin\theta \sin\varphi\omega_{-}^3)^2+(d\theta-\sin\varphi\omega_{-}^2
+\cos\varphi\omega_{-}^3)^2
$$
\be\lb{alfinal}
+(\sin\theta d\varphi
+\sin\theta\omega_{-}^1-\cos\theta \cos\varphi\omega_{-}^2
-\cos\theta \sin\varphi\omega_{-}^3)^2+g_q.
\ee
Let
us introduce the coordinates $u_i$ written in spherical form
$$
u_1=|u| \sin\theta\cos\varphi\cos\phi,
$$
$$
u_2=|u| \sin\theta\cos\varphi\sin\phi,
$$
$$
u_3=|u| \sin\theta\sin\varphi,
$$
$$
u_4=|u| \cos\theta,
$$
Then it is not difficult to check that the
cone $g_8=dr^2+r^2g_7$ being $g_7$ given at (\ref{alfinal}), can be expressed as
\be\lb{explico}
g_s=g|u|^2\overline{g}+f[(du_0-u_i\omega_{-}^i)^2 +(du_i+
u_0\omega_{-}^i + \epsilon_{ijk}u_k\omega_{-}^k)^2].\ee
 The coordinates $u_i$ can be extended to a single quaternion valued coordinate
$$
u=u_0 + u_1 I + u_2 J + u_3 K ,\;\;\;\;\;\; \overline{u}= u_0 - u_1
I - u_2 J - u_3 K.
$$
Here $I, J, K$ denote the unit quaternions, and it follows that
$|du|^2=(du_0)^2+(du_1)^2 +(du_2)^2 + (du_3)^2$. The $Sp(1)\sim
SU(2)$ triplet $\omega_{-}^i$ can be used to define a quaternion
valued one form
$$
\omega_{-}=\omega_{-}^1 I+\omega_{-}^2 J +\omega_{-}^3 K,
$$
and the Kahler triplet $\overline{J}^a$ can be extended to a
quaternion valued two form $\overline{J}= \overline{J}^1 I +
\overline{J}^2 J + \overline{J}^3 K$. The metric (\ref{explico}) can
be expressed in this notation as \be\lb{Swann2} g_8=|u|^2 g_q + |du
+ u \omega_{-}|^2, \ee Under the transformation $u\to G u$ with $G:
M \to SU(2)$ the $SU(2)$ instanton $\omega_{-}$ is gauge transformed
as $\omega_{-}\to G\omega_{-}G^{-1}+ GdG^{-1}$. Therefore the form
$du + \omega_{-}u$ is transformed as
$$
du + u \omega_{-}\rightarrow d(Gu) + (G\omega_{-}G^{-1}+ GdG^{-1})
Gu=G du+ (dG+G\omega_{-}-dG)u=G (du + u \omega_{-}),
$$
and it is seen that $du +  \omega_{-}u$ is a well defined
quaternion-valued one form over the chiral bundle. Associated to the
metric (\ref{Swann2}) we have the quaternion valued two form
\be\lb{quato} \widetilde{\overline{J}}=u\overline{J}\overline{u}+(du
+ u \omega_{-})\wedge \overline{(du + u \omega_{-})}, \ee and it can
be checked that the metric (\ref{explico}) is hermitic with respect to
any of the components of (\ref{quato}). We have that
$$
d\widetilde{\overline{J}}=du\wedge
(\overline{J}+d\omega_{-}-\omega_{-}\wedge
\omega_{-})\overline{u}+u\wedge
(\overline{J}+d\omega_{-}-\omega_{-}\wedge \omega_{-})d\overline{u}
$$
$$+u(d\overline{J}+\omega_{-}\wedge
d\omega_{-}-d\omega_{-}\wedge \omega_{-})\overline{u}.
$$
The first two terms of the last expression are zero due to
(\ref{rela}). Also by introducing (\ref{rela}) into the relation
(\ref{basta}) we obtain that
$$
d\overline{J}+\omega_{-}\wedge d\omega_{-}-d\omega_{-}\wedge
\omega_{-}=0
$$
and therefore the third term is also zero. This means that the
metric (\ref{Swann2}) is hyperkahler with respect to the triplet
$\widetilde{\overline{J}}$ and the holonomy is reduced to
$Sp(2)\subset SU(4)\subset Spin(7)$.

  The hyperkahler metrics (\ref{Swann2}) are indeed well known.
They are the Swann principal $CO(3)$ bundle of co-frames over a
quaternion K\"ahler spaces \cite{Swann}. Hyperkahler quotients of
such metrics by tri-holomorphic isometries are related to quaternion
Kahler quotients of the base spaces.  The hyperkahler condition for
the Swann metric implies that the seven dimensional cone of
(\ref{owner}) is not only Einstein-Sassaki, but tri-Sasaki. The
vector field $\partial_{\phi}$ is the Reeb vector of the tri-Sasaki
metric.
\\

\textit{The self duality of the spin connection}
\\

Although we have found that our example is hyperkahler, it is
instructive to check that the spin connection $\omega_{ab}$ of the
metric (\ref{eyo}) is self-dual. We choose the basis
$$
\overline{e}^{\alpha}=u^{1/2}e^{\alpha},\qquad
\overline{e}^{i}=\frac{\alpha^i}{u^{1/2}},\qquad
\overline{e}^{8}=u^{1/2}(d\tau+H)
$$
being $e^{\alpha}$ a basis for $g_q$ with ${\rm \alpha=1,2,3,4}$ and
${\rm i=1,2,3}$. With the help of the first Cartan equation
$$
d\overline{e}^{m}+\hat{\omega}_{mn}\wedge \overline{e}^{n}=0,
$$
where $m$ an index that can be latin or greek, we obtain the
following components of the spin connection
$$
\hat{\omega}_{ij}=\frac{u^{[i}\alpha^{j]}}{2u^2}-\epsilon_{ijk}\omega^{k}_{-}
+\epsilon_{ijk}\frac{u_k}{u}(d\tau+H),
$$
$$
\hat{\omega}_{\alpha\beta}=\omega_{\alpha\beta}
-\epsilon_{ijk}\frac{u_k}{u^2}(\overline{J}_j)_{\alpha\beta}\alpha^i
-\frac{u_i}{u}(\overline{J}_i)_{\alpha\beta}(d\tau+H),
$$
$$
\hat{\omega}_{\alpha i}=\frac{u^i}{2}
e^{\alpha}-\epsilon_{ijk}\frac{u_k}{u}(\overline{J}_j)_{\alpha\beta}e^{\beta},
$$
$$
\hat{\omega}_{8
i}=\frac{u_i}{u}(d\tau+H)+\epsilon_{ijk}\frac{u_j}{u^2}\alpha^k,
$$
$$
\hat{\omega}_{8
\alpha}=\frac{u_i}{u^2}(\overline{J}_j)_{\alpha\beta}e^{\beta}.
$$
By using that
$2\omega^{i}_{-}=\omega_{\alpha\beta}(\overline{J}_j)_{\alpha\beta}$
and the representation
$$
J^{1}=\left(\begin{array}{cccc}
  0 & -1 & 0 & 0 \\
  1 & 0 & 0 & 0 \\
  0 & 0 & 0 & -1 \\
  0 & 0 & 1 & 0
\end{array}\right),\;\;\;\;
J^{2}=\left(\begin{array}{cccc}
  0 & 0 & -1 & 0 \\
  0 & 0 & 0 & 1 \\
   1 & 0 & 0 & 0 \\
  0 & -1 & 0 & 0
\end{array}\right)
$$
\be\lb{reprodui} J^{3}=J^{1}J^{2}=\left(\begin{array}{cccc}
  0 & 0 & 0 & -1 \\
  0 & 0 & -1 & 0 \\
  0 & 1 & 0 & 0 \\
  1 & 0 & 0 & 0
\end{array}\right),
\ee for the matrix $(\overline{J}_j)_{\alpha\beta}$, it can be
checked that
$$
\hat{\omega}_{81}=-(\hat{\omega}_{23}+\hat{\omega}_{65}+\hat{\omega}_{47}),
\qquad
\hat{\omega}_{82}=-(\hat{\omega}_{31}+\hat{\omega}_{46}+\hat{\omega}_{57}),
$$
$$
\hat{\omega}_{83}=-(\hat{\omega}_{12}+\hat{\omega}_{54}+\hat{\omega}_{67}),
\qquad
\hat{\omega}_{84}=-(\hat{\omega}_{62}+\hat{\omega}_{35}+\hat{\omega}_{71}),
$$
$$
\hat{\omega}_{85}=-(\hat{\omega}_{16}+\hat{\omega}_{43}+\hat{\omega}_{72}),
\qquad
\hat{\omega}_{86}=-(\hat{\omega}_{15}+\hat{\omega}_{24}+\hat{\omega}_{73}),
$$
$$
\hat{\omega}_{87}=-(\hat{\omega}_{14}+\hat{\omega}_{36}+\hat{\omega}_{25}).
$$
These conditions for $\hat{\omega}_{mn}$ can be written more
concisely as
$$
\hat{\omega}_{8i}=-c_{imn}\hat{\omega}_{mn},
$$
and this is equivalent to say that, for the basis $\overline{e}^m$,
the spin connection $\hat{\omega}_{mn}$ is self-dual.

\section{Discussion}

   Along this brief work we have proposed an anzatz for $Spin(7)$ metrics
as an $R$ bundle over closed $G_2$ structures. These $G_2$
structures are $R^3$ bundles over 4-dimensional compact quaternion
K\"ahler spaces. We also have used the fact that the twistor space
of any compact quaternion K\"ahler space is K\"ahler Einstein and
therefore there it possess a six dimensional sympletic form defined
over it. We have imposed the conditions for the reduction of the
holonomy to $Spin(7)$ and we have found a non linear system relating
three unknown functions. We have found a particular solution and the
result was the Swann bundle in eight dimensions, which is
hyperkahler and therefore the holonomy is $Sp(2)\subset Spin(7)$.
Let us recall that the direct sum
$$
g_{11}=g_{1,2}+g_8,
$$
of the Swann metric with the three dimensional flat Minkowski one
$g_{1,2}$ is a solution of the supergravity equations of motion with
all the fields "turned off" except the graviton, and preserving four
supersymmetries after compactification. This solution can be
rewritten in the IIA form
$$
g_{11}= e ^{-\frac{2}{3}\phi} g_{10} + e^{\frac{4}{3}
\phi}(d\tau+H)^2,
$$
being the dilaton $\phi$ defined by $\phi=\frac{3}{4}\log u$. The
reduction along the isometry $\partial_{\tau}$ will give a
background of the form
$$
g_{IIA}= u ^{1/2} g_{1,2} + u ^{-1/2} \widetilde{g}_{7}.
$$
being $\widetilde{g}_7$ given by
$$
\widetilde{g}_7=\frac{(du^1+\omega_{-}^2
u^3)^2}{u^{1/2}}+\frac{(du^2+\omega_{-}^3
u^1)^2}{u^{1/2}}+\frac{(du^3+\omega_{-}^1 u^2)^2}{u^{1/2}} + u^{3/2}
g_q.
$$
The last metric together with the 3-form
$$
\Phi=\frac{1}{u^{3/4}}\alpha_1\wedge \alpha_2\wedge
\alpha_3+u^{5/4}\alpha_i\wedge \overline{J}_i $$ and its dual
\be\lb{cuat} \ast\Phi=u\frac{\epsilon_{ijk}}{2}\alpha_i\wedge
\alpha_j\wedge \overline{J}_k+u^3e_1\wedge e_2\wedge e_3\wedge e_4,
\ee constitute a $G_2$ structure. Therefore and on general grounds
we have that
$$
d\Phi=\tau_o\wedge \ast\Phi+3\tau_1\wedge \Phi+\ast\tau_3,
$$
$$
d\ast\Phi=4\tau_1\wedge \ast\Phi+\tau_2\wedge \Phi
$$
being $\tau_i$ the four torsion classes. We have calculated them for
our case and we have found that $\tau_3=\tau_0=0$ and that the class
$\tau_2$ can be expressed entirely in terms of the sympletic form
$\overline{J}$ for the K\"ahler-Einstein metric. The class $\tau_0$
can be eliminated by a conformal transformation. The expression for
the Swann metric (\ref{owner}) in terms of $\widetilde{g}_7$ is
$$
g_8=u(d\tau+H)^2+\frac{1}{u^{1/2}}\widetilde{g}_7.
$$
 We can therefore paraphrase the
results described in this work by stating that the Swann bundle
defines a conformally closed $G_2$ structure with $\tau_3=\tau_0=0$
by reduction along one isometry (which should not be confused with
an hyperkahler reduction or quotient). It will be interesting to see
if it is possible to find a one parameter deformation of the
solution presented here and to see if the holonomy group obtained is
bigger than $Sp(2)$. In our opinion, this question deserve some
attention.
\\

 {\bf Acknowledgement:} I am benefited with discussions
with G. Giribet, who checked some calculations. Also my
acknowledgments to D. Joyce who pointed out that the example
presented here could be the Swann bundle and to S.Salamon for
presenting me useful literature to compare.
\\

\end{document}